\documentclass{elsart}

\usepackage{color}
\usepackage{times,amssymb, amsmath}

\usepackage{graphicx}

\newcommand{\eps}{\varepsilon}

\begin{document}

\DeclareGraphicsExtensions{.eps}

\begin{frontmatter}

\title{Transition Radiation Spectra of Electrons from 1 to 10
       GeV/c in Regular and Irregular Radiators}

\author[C]{A.~Andronic},
\author[H]{H.~Appelsh{\"a}user},
\author[C]{R.~Bailhache},
\author[I]{C.~Baumann},
\author[C]{P.~Braun-Munzinger},
\author[I]{D.~Bucher},
\author[C]{O.~Busch},
\author[B]{V.~C{\u a}t{\u a}nescu},
\author[D]{S.~Chernenko},
\author[A]{P.~Christakoglou},
\author[D]{O.~Fateev},
\author[F]{S.~Freuen},
\author[C]{C.~Garabatos},
\author[I]{H.~Gottschlag},
\author[J]{T.~Gunji},
\author[J]{H.~Hamagaki},
\author[F]{N.~Herrmann},
\author[I]{M.~Hoppe},
\author[E]{V.~Lindenstruth},
\author[C]{C.~Lippmann\corauthref{cor1}}
\corauth[cor1]{Corresponding author},
\ead{C.Lippmann@gsi.de}
\ead[url]{http://www-linux.gsi.de/{\~{}}lippmann}
\author[J]{Y.~Morino},
\author[D]{Yu.~Panebratsev},
\author[A]{A.~Petridis},
\author[B]{M.~Petrovici},
\author[F]{I.~Rusanov},
\author[C]{A.~Sandoval},
\author[J]{S.~Saito},
\author[F]{R.~Schicker},
\author[F]{H.K.~Soltveit},
\author[F]{J.~Stachel},
\author[C]{H.~Stelzer},
\author[A]{M.~Vassiliou},
\author[F]{B.~Vulpescu},
\author[I]{J.P.~Wessels},
\author[I]{A.~Wilk},
\author[D]{V.~Yurevich},
\author[D]{Yu.~Zanevsky}

for the ALICE collaboration.

\address[C]{GSI Darmstadt (Germany)}
\address[H]{Institut f{\"u}r Kernphysik, University of Frankfurt (Germany)}
\address[I]{Institut f{\"u}r Kernphysik, University of M{\"u}nster (Germany)}
\address[B]{NIPNE Bucharest (Romania)}
\address[D]{JINR Dubna (Russia)}
\address[A]{University of Athens (Greece)}
\address[F]{Physikalisches Institut, University of Heidelberg (Germany)}
\address[J]{University of Tokyo (Japan)}
\address[E]{Kirchhoff-Institut f{\"u}r Physik, University of Heidelberg (Germany)}


\begin{abstract}

  We present measurements of the spectral distribution of transition radiation
  generated by electrons of momentum 1 to 10\,GeV/c in different radiator types.
  We investigate periodic foil radiators and irregular foam and fiber materials.
  The transition radiation photons are detected by prototypes of the drift
  chambers to be used in the Transition Radiation Detector (TRD) of the ALICE
  experiment at CERN, which are filled with a Xe, CO$_2$ (15\,\%) mixture. The
  measurements are compared to simulations in order to enhance the
  quantitative understanding of transition radiation production, in particular
  the momentum dependence of the transition radiation yield.

\end{abstract}


\end{frontmatter}

\section{Introduction}
\label{Intro}

The ALICE Transition Radiation Detector (TRD)\,\cite{TRD,posres} is a
large area (750\,m$^2$,
1.18 million readout channels) device that offers precise tracking  and
electron identification and -- combining these two capabilities --
a fast trigger on high-$p_t$ electrons and jets. The readout chambers
are drift chambers with a 3\,cm drift region
and a 0.7\,cm amplification region, separated by a cathode wire grid.
They are read out by cathode pads of varying sizes via charge sensitive
preamplifiers/shapers (PASA). The maximum drift time is about 2\,$\mu$s
and the induced signal is sampled on all channels at 10\,MHz to record
the time evolution of the signal\,\cite{signal,gain}. A {\it sandwich}
radiator is placed in front of each gas volume, which is a box structure
of foam ({\it Rohacell} HF71, 8\,mm), reinforced by a carbon fiber
laminate coating on both sides and containing seven layers
of polypropylene fiber sheets ({\it Freudenberg LRP375BK}, 5\,mm each).
A drift cathode made of 12\,$\mu$m aluminized mylar is laminated onto
the carbon fiber. This design was chosen in order to find a compromise
between maximum transition radiation yield and a solid construction for
our large area detectors.

Transition radiation (TR) is emitted by particles traversing the
radiator with a velocity larger than a certain threshold\,\cite{TR},
which corresponds to a Lorentz factor of $\gamma \approx 1000$. The
produced TR photons have energies in the X-ray range, with the bulk
of the spectrum from 1 to 30\,keV. In the drift chambers
of the ALICE TRD a gas mixture of Xe and CO$_2$ (15\,\%) is used
to provide efficient absorption of these photons. To discriminate
electrons from the large background of pions in the momentum range
of interest (1 to 10\,GeV/c), two characteristic phenomena
are used \cite{pioeff}:

\begin{itemize}
\item[i)] The ionization energy loss\,\cite{dedx} is larger for
  electrons than for pions, since electrons are at the plateau of
  ionization energy loss, while pions are minimum ionizing or in
  the regime of the relativistic rise.
\item[ii)] Only electrons reach a velocity that exceeds the TR
  production threshold.
\end{itemize}

In the ALICE TRD, as in any other TRD, TR is superimposed on the ionization
energy loss, due to the small angle of emission ($\backsim 1/\gamma$) with
respect to the emitting particles trajectory. This explains why the
existing measurements of TR spectra\,\cite{TRspec,Fabjan,ch1,ch2} are
scarce. In this publication, we investigate specific features of TR, in
particular its spectral distributions and momentum dependence produced in
the ALICE TRD {\it sandwich} radiators. By improving the experimental setup
with respect to a similar earlier measurement\,\cite{TRspec}, we are
able to cover the full momentum range from 1 to 10\,GeV/c. In addition,
we also investigate different irregular radiators and two regular foil
radiators.

\section{Experimental Setup and Analysis Method}
\label{ExpSetup}

\begin{figure}[htp]
  \begin{center}
    \includegraphics[width=10cm]{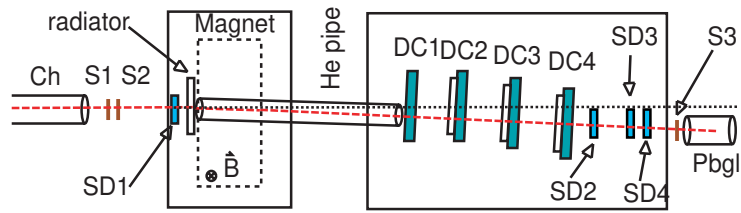}
    \caption{Setup of the beam test. A radiator is placed in front of
      the dipole magnet, which separates the beam from the TR photons
      produced in the radiator. A pipe filled with helium is used to
      minimize the absorption of the TR photons on the $\sim$1.6\,m long
      path to the drift chamber (DC1). Three other
      drift chambers (DC2-DC4) have a radiator in front and are used for
      reference and for d$E$/d$x$ measurements. Electrons can be selected
      by coincident thresholds on a \u{C}erenkov detector (Ch) and a
      lead-glass calorimeter
      (Pbgl). Three scintillating detectors (S1-S3) are used for
      triggering and four silicon detectors (SD1-SD4) for position
      reference.}
    \label{setup}
   \hspace{5mm}
  \end{center}
\end{figure}

\begin{figure}[t]
  \begin{center}
    \includegraphics[width=8cm]{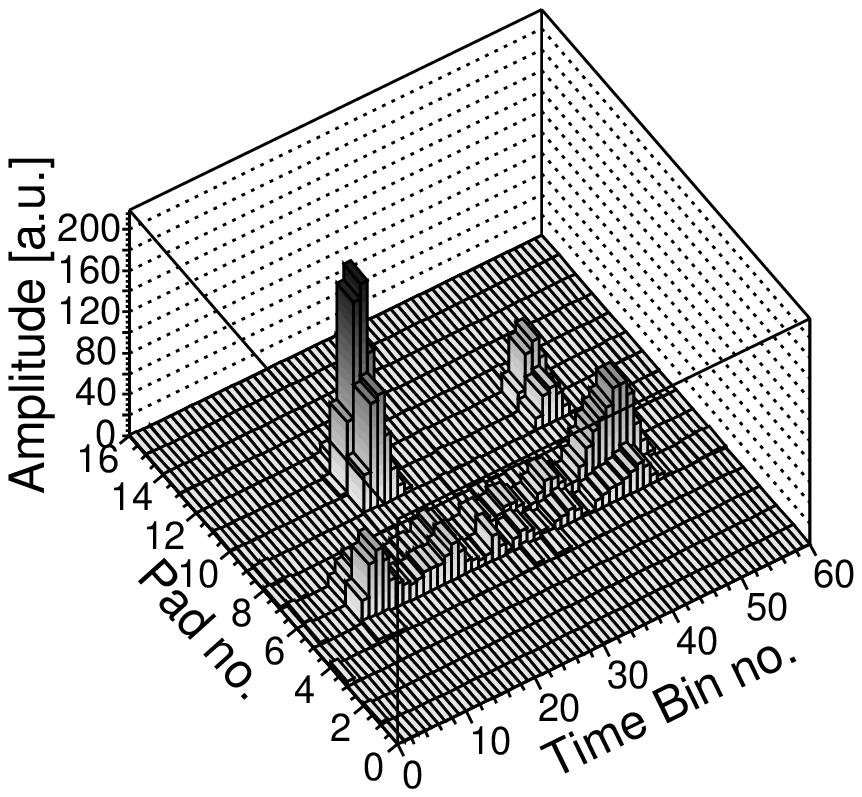}
    \caption{Pulse height versus drift time on sixteen adjacent cathode
    pads for an example event. The maximum drift time at the electron
    drift velocity of 1.5\,cm/$\mu$s is about 2\,$\mu$s, thus at the
    employed sampling frequency of 20\,MHz one time bin corresponds to
    about 0.75\,mm in the drift direction perpendicular to the wire planes.
    The time zero is arbitrarily shifted by 0.5\,$\mu$s
    to facilitate a simultaneous measurement of the baseline and of noise.
    The baseline is subtracted in the shown event. Here, the ionization
    signal by a beam electron and two well separated photon clusters
    are clearly visible.}
    \label{event}
   \hspace{5mm}
  \end{center}
\end{figure}

The measurements were carried out at the T9 secondary beam line at the CERN
PS accelerator. The beam consisted of electrons and pions with momenta of
1 to 10\,GeV/c. The method to study TR is similar to that described in
\cite{Fabjan};
a sketch of our setup is shown in Fig. \ref{setup}. A dipole magnet is used
to deflect the beam in order to spatially separate the TR photons from the
path of the beam (see Fig. \ref{event}). A 1.6\,m long pipe filled with
helium is used to minimize photon absorption.
The photons also have to cross the two entrance windows to the helium
pipe (20\,$\mu$m mylar foil each), a small amount of air ($\approx$1\,mm
on both sides) and the entrance window (drift cathode) of the prototype
drift chamber (20\,$\mu$m mylar foil with aluminium sputtering,
$\approx$1\,$\mu$m). The length of the magnet is approximately 0.5\,m.
The radius of a beam particle is kept approximately constant (33\,m) for
the different momenta by adjusting the strength of the magnetic field
up to about 1\,T at 10\,GeV/c.

\begin{figure}[tp]
  \begin{center}
    \includegraphics[width=8cm]{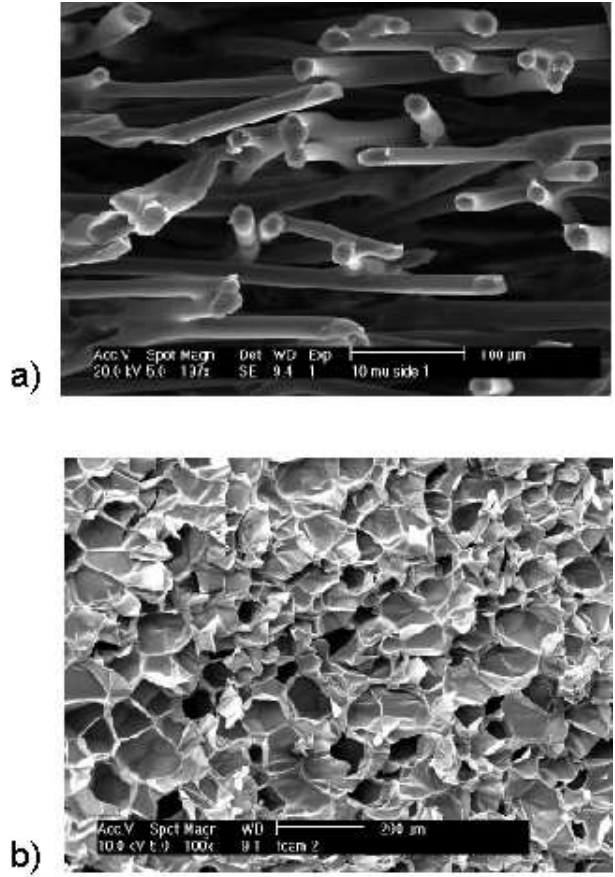}
    \caption{SEM pictures of the irregular radiator materials:
    Polypropylene fibers ({\it Freudenberg LRP375BK}) (a) and
    {\it Rohacell} foam (HF71) (b).}
    \label{radiators}
   \hspace{5mm}
  \end{center}
\end{figure}

The parameters of all radiators investigated in this work are given
in table \ref{radiat}. The {\it foam} and {\it fiber} radiators are each
made of the same materials which are components of the ALICE TRD
{\it sandwich} (see Fig. \ref{radiators}). The dummy radiator is made of
plexiglas.

\begin{table}[ht]
  \begin{center}
    \begin{tabular}{|c||c|c|c|c|c|c|}
      \hline
      & thickness & $N_f$ & $d_1$ & $d_2$
      & density & $X/X_0$\\[-2.5mm]
      & [cm] & [ ] & [$\mu$m] & [$\mu$m] & [g/cm$^3$] & [\%] \\[-1mm]
      \hline
      \hline
      {\it Sandwich} & 4.8 & (115) & (13) & (400) & - & 0.87 \\[-1mm]
      \hline
      {\it Fiber} & 4.0 & (115) & (13) & (400) & 0.074 & 0.65 \\[-1mm]
      \hline
      {\it Foam} & 4.2 & - & - & - & 0.075 & 0.78 \\[-1mm]
      \hline
      {\it Reg1} & 6.24 & 120 & 20 & 500 & - & 0.61 \\[-1mm]
      \hline
      {\it Reg2} & 5.94 & 220 & 20 & 250 & - & 1.37 \\[-1mm]
      \hline
      {\it Dummy} & 0.4 & - & - & - & 1.18 & 1.2\\
      \hline
    \end{tabular}
    \caption{Properties of the various radiators investigated. $N_f$ and
    $d_1$ are the number and thickness of the foils, $d_2$ is the spacing.
    For the {\it sandwich} and {\it fibers} radiator the numbers used as
    parameters in the simulation are given in parenthesis.}
     \label{radiat}
  \end{center}
\end{table}

We use four identical prototype drift chambers with a geometry
similar to that of the final TRD chambers, but with a smaller active area
(25\,$\times$\,32\,cm$^2$). The dimensions of the pads are
0.75\,$\times$\,8\,cm$^2$. We use the final version of the ALICE TRD
PASA with an on-detector noise of about 1000 electrons (r.m.s.). The
drift chambers are read out with a Flash ADC system at 20\,MHz, which is
twice the nominal sampling rate of the TRD, to increase the position
resolution in the drift direction. The high voltage at the anode wires is
adjusted to provide a relatively low gas gain (around 4000) to minimize space
charge effects\,\cite{sc}. A few presamples are taken for each channel
in order to extract information on the baseline and noise.
The signal generated by the ionization energy loss and by the TR
photons is measured simultaneously on a row of 16 readout pads. An
example event is shown in Fig. \ref{event}. For each event a TR
cluster search is performed. We are able to provide a separation
from the beam of more than three pads at all momenta, which avoids
a contamination of the measured TR energy with ionization energy.
The obtained charge spectra are calibrated using the 5.96\,keV
line of $^{55}$Fe\,\cite{dedx} and by comparing the most probable pion
energy loss in the TRD chambers to earlier measurements and to
simulations (including corrections for ambient variations).

\section{Transition Radiation and Detector Performance Simulations}
\label{Sim}

In general, all ALICE related simulations are carried out using
AliRoot\,\cite{Aliroot}, the ALICE software package. The interaction of
the charged particles with the detector materials and their energy loss
are currently simulated using Geant 3.21\,\cite{Geant}. In principle the
employed ``Virtual Monte Carlo'' would allow one to run also different
simulation engines (Geant 4\,\cite{Geant4} or Fluka\,\cite{Fluka})
without changing the user code containing the input and output format
nor the geometry and detector response definition.

For the simulation of the performance of the TRD detector system a
quantitative understanding of TR is indispensable. Since the
production of TR is not included in Geant 3, we have
explicitly added it to AliRoot. We use an approximation for the TR
yield of a regular stack of foils with fixed thickness, including
absorption\,\cite{Fabjan}, which we tune to reproduce the measured
performance of the irregular ALICE TRD radiators\,\cite{pioeff,TRspec}.
Geant 4 is in principle able to simulate the TR emitted by regular
and irregular structures\,\cite{G4}. A satisfactory comparison to
our measured data has however not been accomplished yet.

\begin{figure}[ht]
  \begin{center}
    \includegraphics[width=8cm]{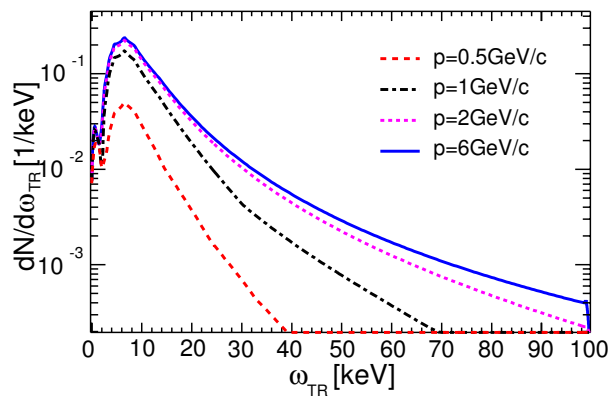}
    \caption{Calculated TR spectra at the end of a regular radiator
    with $N_f = 220$, $d_1 = 20\,\mu$m and $d_2 = 250\,\mu$m ({\it Reg2})
    for electrons at four momenta. We find
    $\langle N_{TR}^{prod}\rangle$=0.46, 1.68, 2.41 and 2.63 for electrons
    at $p=0.5$, 1, 2 and 6\,GeV/c, respectively. The distributions are
    obtained using the formalism of Ref. \cite{Fabjan}.}
    \label{SpecProd}
   \hspace{5mm}
  \end{center}
\end{figure}

A formula for the energy spectrum of the TR photons emitted by a highly
relativistic electron passing through a radiator composed of $N_f$ foils
of thickness $d_1$, spaced periodically by gaps of width $d_2$, has been
given in\,\cite{Fabjan}. Some example spectra of the energy of TR photons
exiting a regular radiator calculated using that formula are shown in
Fig. \ref{SpecProd}. The onset of TR production takes place around
0.5\,GeV/c and the predicted TR yield essentially saturates above
2\,GeV/c. The average number $\langle N_{TR}^{prod}\rangle$
of produced TR photons per electron event is given by the integral
of these spectra. To simulate the emission and detection of TR
photons we proceed in three steps:

\begin{enumerate}
\item Draw a number from a Poisson distribution with a mean given by
  $\langle N_{TR}^{prod}\rangle$,
\item calculate the energies of these produced photons by drawing a random
  number from the differential energy spectrum (Fig. \ref{SpecProd}) and
\item propagate these photons through the experimental setup.
\end{enumerate}

All materials crossed by the photons, e.g. gas volumes and foil
windows are taken into account by including the energy dependent
absorption lengths of the given materials in the simulation. We use
tabulated X-ray mass attenuation coefficients from Ref.\,\cite{xray}.
The detector energy resolution of 32\,\% ($\sigma$), as measured
with a $^{55}$Fe source, is folded in.

\section{Results}
\label{Results}

\subsection{Cluster Number Distribution}

\begin{figure}[ht]
  \begin{center}
    \includegraphics[width=8cm]{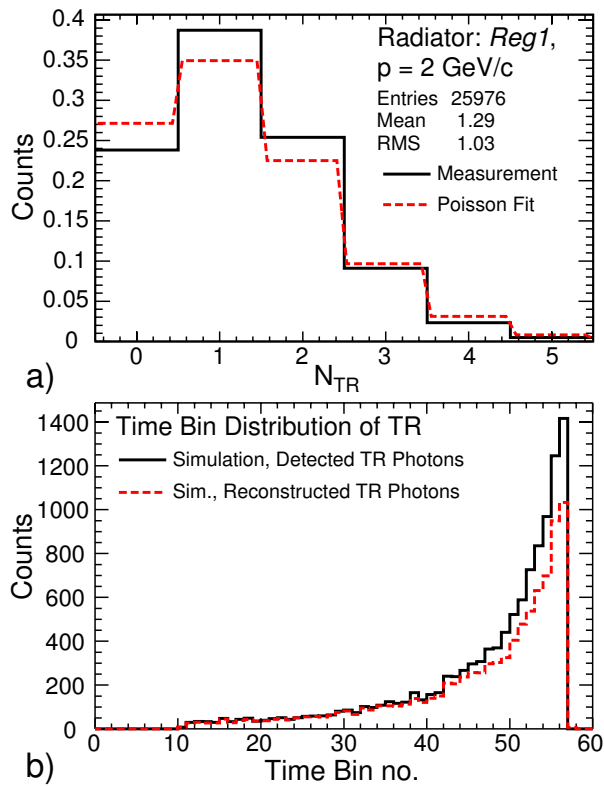}
    \caption{a) Photon number distribution for 2\,GeV/c beam momentum
    for the radiator {\it Reg1} together with a Poisson fit.
    b) Simulated time bin distributions for the absorption of TR photons.}
    \label{tr-reg2}
   \hspace{5mm}
  \end{center}
\end{figure}

In Fig. \ref{tr-reg2}a we present the distribution of the detected
number of photons per incident electron for 2\,GeV/c electrons for
the radiator {\it Reg1}. The shape of the distribution can
be approximated by a Poissonian, indicated by the dashed histogram.
The minimum time interval between two TR photons resolved by the
algorithm is given by two time bins (100\,ns), corresponding to
about 1.5\,mm. This can lead to a considerable amount of events
with overlapping clusters. Fig. \ref{tr-reg2}b shows for simulated
events the time bin distribution of the detected TR clusters and of
the clusters reconstructed by the TR search algorithm. Close to the
drift cathode (for large time bin numbers) the density of the TR
hits is largest and, as a consequence, the probability to misidentify
two close photon events to be a single TR photon one is highest in
this region.

\subsection{Data as a Function of Momentum}\label{ResMom1}

\begin{figure}[ht]
  \begin{center}
    \includegraphics[width=8cm]{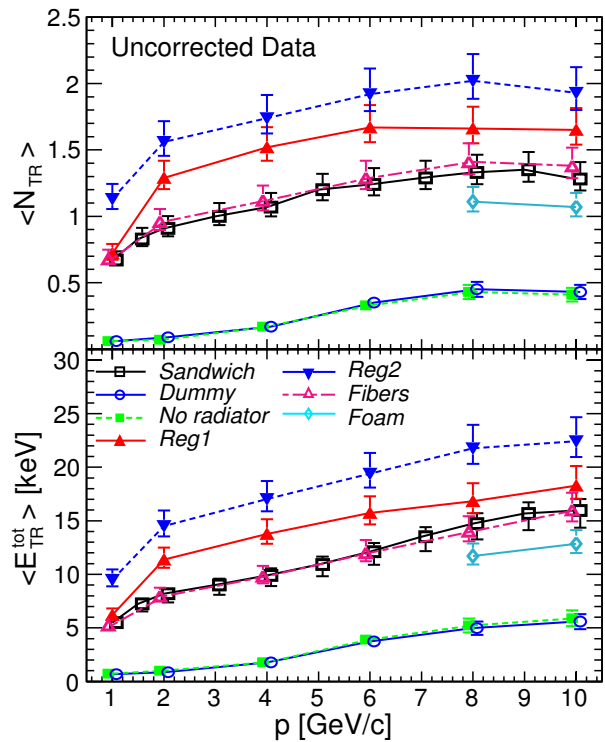}
    \caption{Measured momentum dependence of the mean number of
    reconstructed TR photons (upper panel) and of the mean total
    TR energy measured as a function of momentum (lower panel).}
    \label{tr-mom1}
   \hspace{5mm}
  \end{center}
\end{figure}

While for our earlier measurement of TR spectra the range in momenta
covered was restricted by the possible range of the magnetic field
and by the experimental configuration\,\cite{TRspec}, we are now able
to present measurements over a broad momentum range from 1 to 10\,GeV/c.
The (uncorrected) results for all radiators are shown in Fig.
\ref{tr-mom1}. Here $\langle N_{TR}\rangle$ denotes the mean number of
detected TR clusters and $\langle E_{TR}^{tot}\rangle$ is the mean
energy deposited in the detector by TR by one electron. The energy
threshold of the TR cluster search algorithm is about 0.1\,keV.
For all radiators, a systematic increase of the photon yield as a
function of momentum is observed. The two regular radiators show
the best performance, followed by the {\it fiber} radiator and the
{\it sandwich}. For the {\it foam} radiator only two measurements
at 8 and 10\,GeV/c are available; the photon yield of this material is
somewhat lower than for the other cases. With the {\it dummy}
radiator and without radiator we observe a considerable amount of
photons in the drift chamber. This photon background is investigated
in the following section.

\subsection{Photon Background}\label{Results2}

TR from our radiators is not the only possible source of photons detected
in the drift chamber. One could suspect TR from different sources (material
in the beam line), bremsstrahlung from material in the beam (or from the
radiators) and synchrotron radiation from the electrons in the magnetic
field. The influence of bremsstrahlung created in the radiators can be
investigated by looking at the data for the dummy radiator (since its
radiation length is comparable to the radiation length of the radiators,
see Table \ref{radiat}) and comparing it to the data without radiator
(see Fig. \ref{tr-mom1}). Since the photon yield is very similar, we can
exclude that the influence of bremsstrahlung from the radiators is
significant.

\begin{figure}[ht]
  \begin{center}
    \includegraphics[width=8cm]{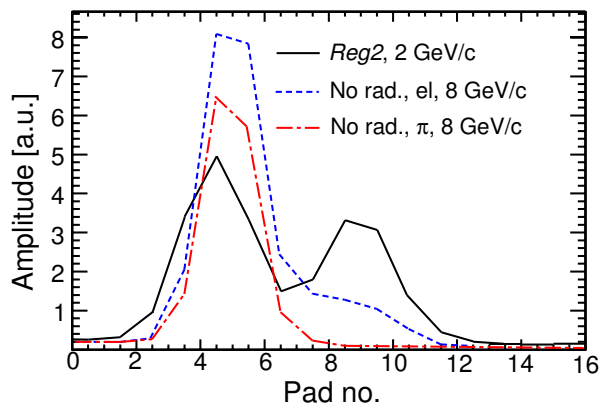}
    \caption{Example distributions of the average signal amplitudes on
    the 16 readout pads. We compare a run with the {\it Reg2} radiator
    to a run without radiator. The peak around pad numbers 4 and 5
    corresponds to the position of the beam. Distributions to the
    right of these peaks indicate charge deposit by photons (TR and/or
    SR).}
    \label{srpad}
   \hspace{5mm}
  \end{center}
\end{figure}

While the contribution of TR from other material in the beam
line is expected to be very small, the production of synchrotron
radiation (SR) can be significant. This explanation is supported
by the distribution of the average signal amplitudes across the pads
shown in Fig. \ref{srpad}. For pions the only contribution is the
ionization energy deposit. For electrons, the contribution of TR is
well separated from the ionization energy deposit. On the other hand,
the distribution of deposited energy for electrons at high momenta
and without a radiator shows a shape that corresponds to what is
expected from SR. In this case SR photons are emitted continuously
along the curved electron tracks and their absorption with respect
to the beam position in the detector is expected to be uniform up
to a certain distance from the particle track.

\subsection{Simulation of Synchrotron Radiation}
\label{Sim2}

Synchrotron radiation (SR) occurs when charged particles move on a
curved path. Since the energy radiated by the particle is proportional
to $\gamma^4$, it can be produced with a sizeable yield in our momentum
range by electrons only. In the following treatment, the magnetic field
is assumed to be constant and uniform, neglecting possible fringe field
effects. Then the energy radiated by an ultra-relativistic electron
along a trajectory of length $L$ is \cite{G4UG}

\begin{equation}
  \Delta E_{SR} \ = \ \frac{2}{3}\,\frac{L}{R^2}\,
  \frac{e^2}{4\pi\eps_0}\ \beta\ \gamma^4 \quad
  \mbox{(for}\quad L \ll R\,\mbox{)}\ .
  \label{rad_energy}
\end{equation}

The velocity $\beta$ of the particle is measured in units of the speed
of light. We are interested in the spectral distribution of the radiated
photons $\frac{dN}{d\omega_{SR}}$. This can be expressed in terms of the
mean energy loss spectrum

\begin{equation}
  \frac{dN}{d\omega_{SR}} \ = \ \frac{\sqrt{3}}{2\,\pi}\,
  \alpha\,\frac{L\,\gamma}{R}\,
  \frac{1}{\omega_C}\int\limits_{\omega_{SR}/\omega_C}^{\infty}K_{5/3}(\eta)\
  d\eta \ ,
  \label{srspectrum}
\end{equation}

\begin{figure}[ht]
  \begin{center}
    \includegraphics[width=8cm]{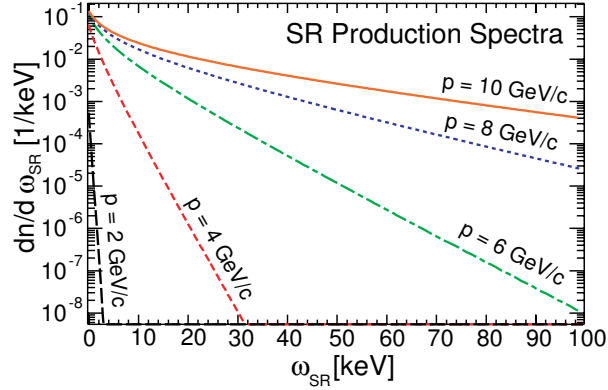}
    \caption{Synchrotron Radiation (SR) production spectra as calculated
      using Eq. \ref{srspectrum} for electrons at five momenta, $L=0.5$\,m,
      $R=33$\,m.}
    \label{SpecProdSR}
   \hspace{5mm}
  \end{center}
\end{figure}

with $\omega_{SR}$ the synchrotron photon energy, $\alpha$ the fine
structure constant and $K_{5/3}$ the MacDonald function\footnote{The
  modified Bessel function of the second kind.}. The characteristic
energy $\omega_C$ of the SR is given by
$1.5\,\frac{\beta\,\hbar\,c}{R}\,\gamma^3$. Examples of SR energy
spectra are shown in Fig. \ref{SpecProdSR}. The mean number of
produced SR photons at all energies is the integral of the spectral
distribution (Eq. \ref{srspectrum}), which can be approximated as

\begin{equation}
  \langle N^{prod}_{SR}\rangle \ \approx \ 10^{-2}\ \frac{L\,\gamma}{R}\ .
  \label{mean_n}
\end{equation}

\begin{figure}[ht]
  \begin{center}
    \includegraphics[width=8cm]{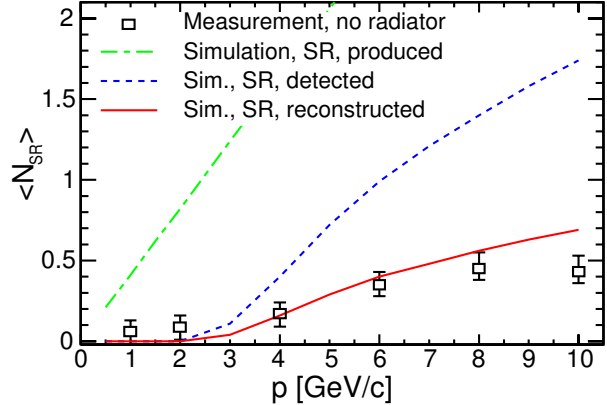}
    \caption{Measured number of detected photons without radiator
    as a function of the beam momentum together with simulations for
    synchrotron radiation (SR).}
    \label{sr-mom}
   \hspace{5mm}
  \end{center}
\end{figure}

\begin{figure}[ht]
  \begin{center}
    \includegraphics[width=8cm]{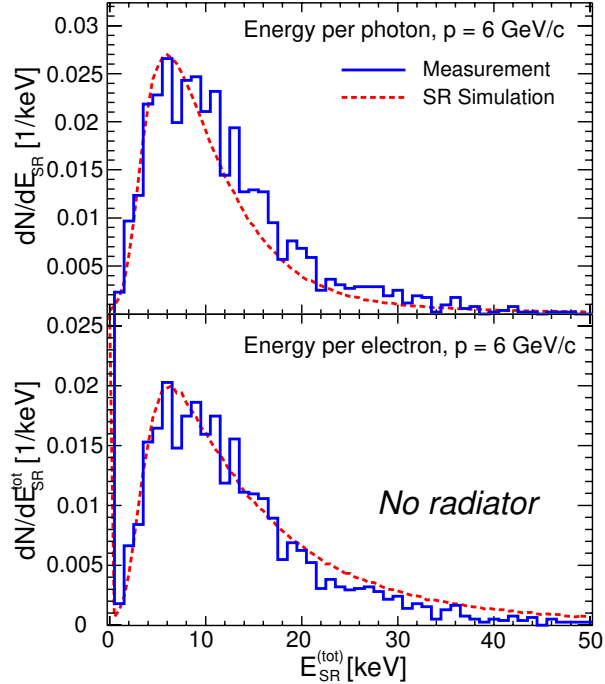}
    \caption{Measured photon energy spectra without radiator together
    with simulated SR spectra at 6\,GeV/c. The upper panel shows the
    distribution of the energy for single photons. Here the
    simulated spectrum is scaled down by a factor of 0.4 to account for
    the photon detection efficiency of the algorithm due to
    the overlap of the photon absorption with the energy deposit from
    the beam particles. The lower panel shows the distribution of the
    total SR energy. The entries in the bin at 0\,keV (off scale)
    correspond to events when no photon was detected.}
    \label{sr-spec}
   \hspace{5mm}
  \end{center}
\end{figure}

Fig. \ref{sr-mom} shows the measured photon background from Fig.
\ref{tr-mom1} (without radiator) together with SR simulations. The mean
number of SR photons produced per electron along the curved electron
track is about 0.4 at 1\,GeV/c and rises to around 4 at 10\,GeV/c.
However, most low energy photons are absorbed in the material before
the drift chamber volume, while at higher momenta some high energy
photons ($\gtrsim$30\,keV) exit the drift chamber undetected.
As a consequence, we find that no photons are detected up to
2\,GeV/c electron momentum and about 1.7 at 10\,GeV/c. The
cluster search algorithm is not able to reconstruct photons that
are deposited very close to the beam, where a large part of the SR
photons are expected to be deposited. By running the cluster search
algorithm on simulated SR data we observe a reconstruction
efficiency of around 40\,\%. Taking this into account, the
agreement with the measurement is very good and we conclude that we
observe a background of synchrotron radiation in our measurements.
This background can be subtracted when investigating mean values.
In addition, in the momentum region up to 2\,GeV/c the contribution of
SR is negligible, and as a consequence our TR energy spectra for
those momenta are free from SR. Fig. \ref{sr-spec} shows measured
photon energy spectra without radiator together with simulated SR
spectra for electrons of 6\,GeV/c. The agreement is quite good.

\subsection{TR Spectra}\label{ResSpectra}

In this section we compare the measured spectral distributions of TR for
2\,GeV/c electrons to simulations. The low energy threshold of the TR search
algorithm (0.1\,keV) and the dynamic range of the employed Flash ADC
system lead to meaningful data in the energy range of interest.

\begin{figure}[p]
  \begin{center}
    \includegraphics[width=14cm]{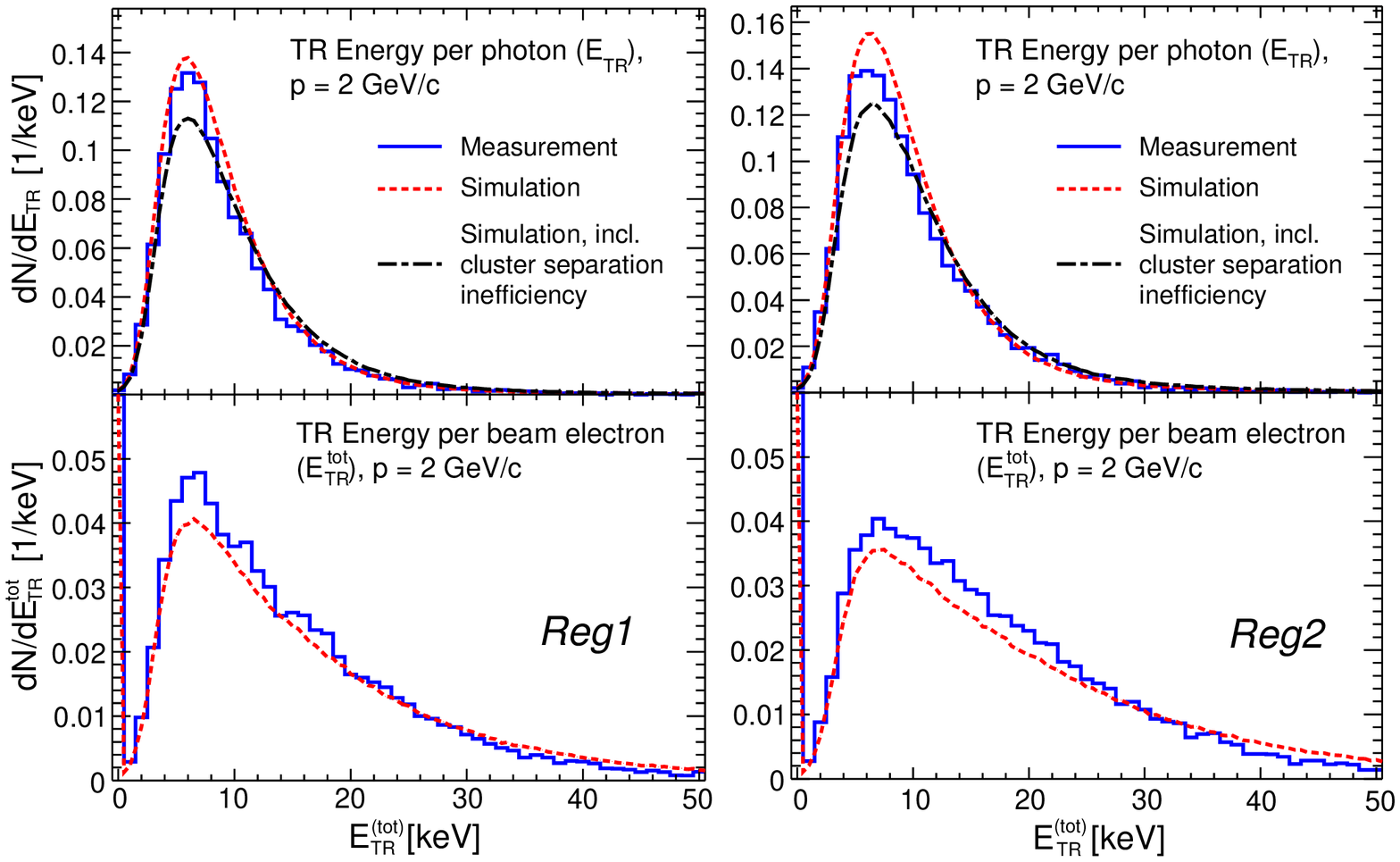}
    \caption{Measured and simulated TR spectra at 2\,GeV/c for the
    two regular foil radiators {\it Reg1} (left panels) and {\it Reg2}
    (right panels). The upper panels show the distribution of the
    energy for single TR photons. The lower panels show the
    distribution of the total TR energy detected per incident beam
    electron. The entries at 0\,keV (off scale) correspond to events
    when no photon was detected.}
    \label{tr-spec-reg}
   \hspace{5mm}
  \end{center}
\end{figure}

In Fig. \ref{tr-spec-reg} we show the measured distributions of TR
energy per photon ($E_{TR}$) and per incident electron (total TR
energy, $E_{TR}^{tot}$) for the two regular radiators at 2\,GeV/c
together with simulations. The simulated spectra are obtained
using the procedure described in section \ref{Sim}, where the
values of $N_f$, $d_1$ and $d_2$ are fixed by the geometries of
the radiators (see table \ref{radiat}). If a correction for the
cluster separation inefficiency is performed the agreement is
good. However, the simulated spectra are somewhat harder in
both cases.

The observable $E_{TR}^{tot}$ is not affected by the TR search
algorithm. Events with $E_{TR}^{tot} = 0$ are mainly due to
events with no TR photon ($N_{TR}=0$). Again, we observe that
the simulated spectra are harder, especially for the radiator
{\it Reg2}. This discrepancy can not be explained by slight
variations of the foil number, thickness or spacing.

\begin{figure}[p]
  \begin{center}
    \includegraphics[width=14cm]{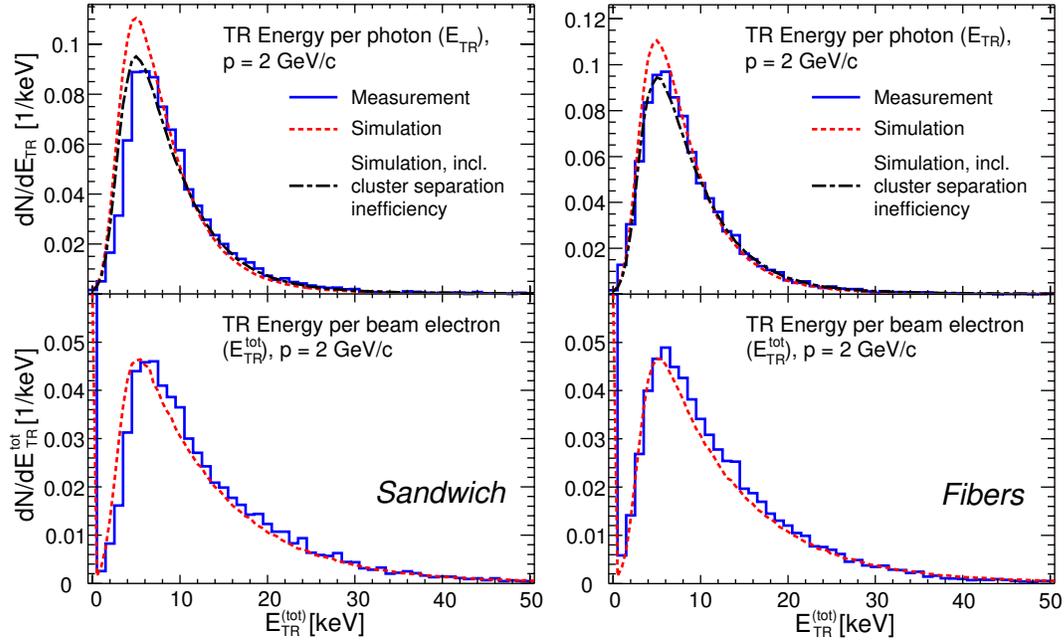}
    \caption{Measured and simulated TR spectra at 2\,GeV/c for the
    ALICE TRD {\it sandwich} radiator (left panels) and for the {\it fiber}
    radiator (right panels).}
    \label{tr-spec-sandw}
   \hspace{5mm}
  \end{center}
\end{figure}

Measured and simulated TR spectra for the ALICE TRD {\it sandwich}
radiator and for the {\it fiber} radiator at 2\,GeV/c are shown
in Fig. \ref{tr-spec-sandw}. The spectra for the two radiators
are very similar, which is surprising, since the {\it sandwich} radiator
also contains the carbon fiber and mylar layers and some
amount of epoxy. In the simulation we use the theory described in
section \ref{Sim}. Since the two radiators are not regular foil
structures, the parameters $N_f$ , $d_1$ and $d_2$ can only
reflect typical dimensions of the radiator materials (see Fig.
\ref{radiators}), but are not unambiguously determined. Our set
of parameters ($N_f = 115$, $d_1 = 13\,\mu$m and $d_2 = 400\,\mu$m)
was chosen to best reproduce our measurements, taking into account
also the influence for the limited cluster separation efficiency.
This set of parameters leads to a good agreement with the data
for both radiator types.

\subsection{Data Corrected for Synchrotron Radiation}
\label{ResCorr}

\begin{figure}[ht]
  \begin{center}
    \includegraphics[width=14cm]{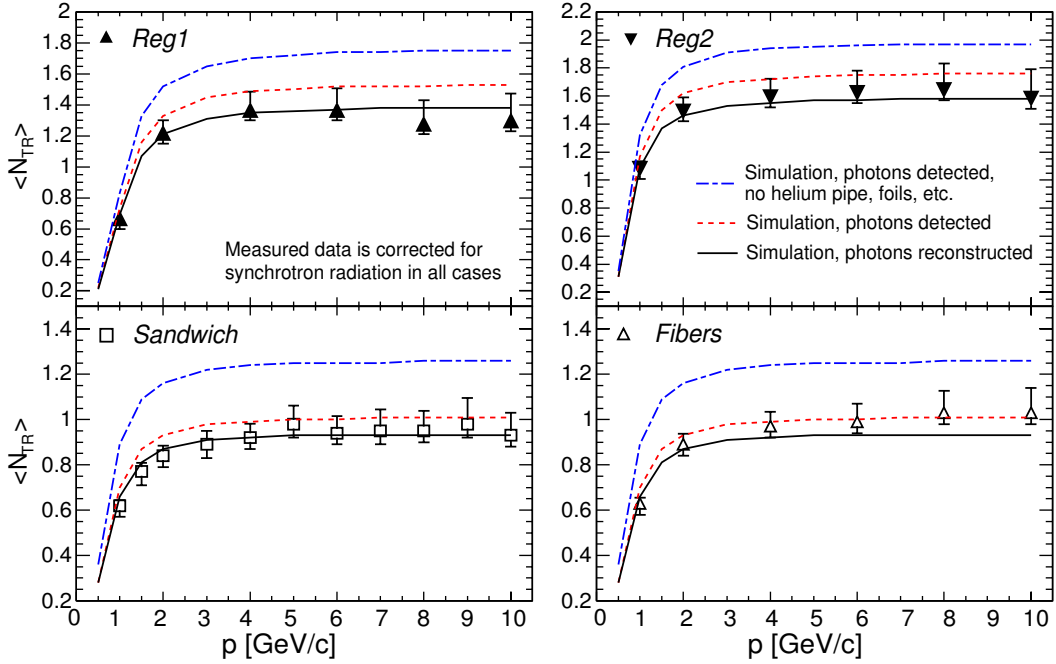}
    \caption{Measured and simulated dependence of the mean number of
    detected TR photons $\langle N_{TR}\rangle$ on the momentum for the
    two regular radiators and for the {\it sandwich} and {\it fiber}
    radiators. The measured data are corrected for SR; the simulated data
    show the actual detected mean number of TR photons with and without
    the helium pipe and its entrance windows included in the simulation
    and the mean number of photons reconstructed by the TR cluster
    search algorithm.}
    \label{ntr-mom-corr}
   \hspace{5mm}
  \end{center}
\end{figure}

We perform the correction of the measured mean values of the TR yield
for the photon background due to SR by subtracting the mean values
obtained without radiator. Since we performed
measurements for the {\it sandwich} radiator at more momenta than
without radiator, the data at these momenta are obtained by the
subtracting interpolated values for the SR background. Fig.
\ref{ntr-mom-corr} shows the momentum dependence of the mean number
of TR photons $\langle N_{TR}\rangle$ for the two regular radiators,
for the {\it sandwich} radiator and for the {\it fiber} radiator. We
find that for
all radiators up to roughly 0.2 photons are absorbed in the helium
pipe, the entrance windows and the air in between. We also include
in the simulation the full TR cluster search algorithm that was
used on the measured data. This leads to a good reproduction of the
measured data for the two regular radiators, in contrast to
earlier findings, where the simulations overestimated the data, in
particular for Xe-based detectors\,\cite{ch1,ch2}. As already
mentioned, in the case of the irregular radiators the parameters for
the simulation ($N_f$, $d_1$ and $d_2$) were tuned to best reproduce
the measurements.

\begin{figure}[ht]
  \begin{center}
    \includegraphics[width=14cm]{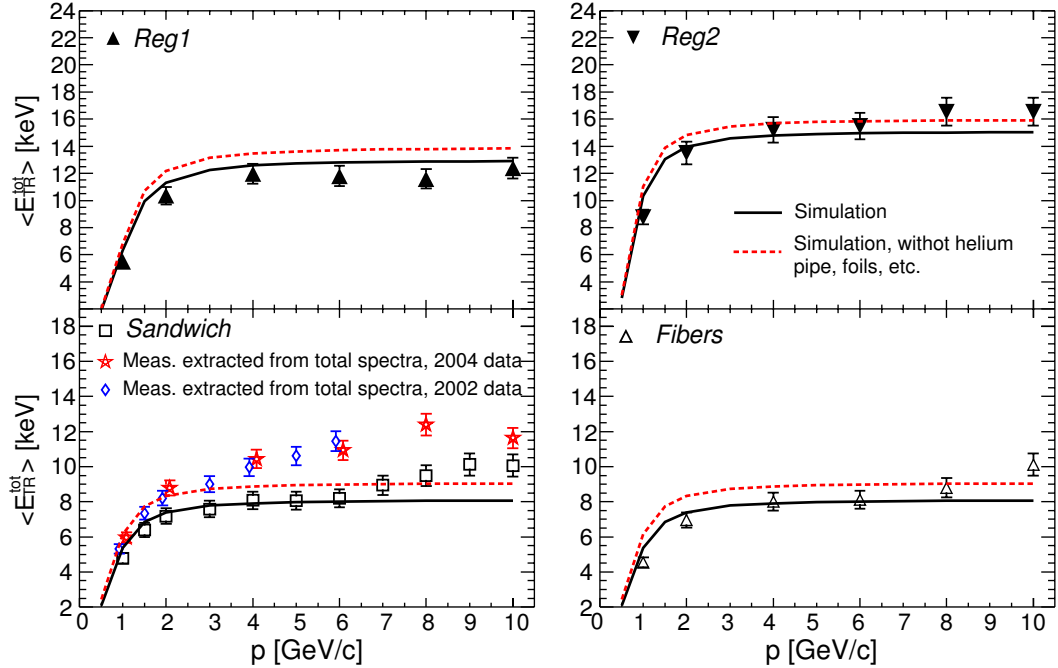}
    \caption{Measured and simulated total TR energy
    $\langle E_{TR}^{tot}\rangle = \langle E_{TR}\rangle\langle N_{TR}\rangle$
    as a function of momentum for the two regular foil radiators and for
    the {\it sandwich} and {\it fiber} radiators. For the {\it sandwich}
    radiator we also include mean values extracted from the total energy
    loss spectra.}
    \label{etrntr-mom}
   \hspace{5mm}
  \end{center}
\end{figure}

Fig. \ref{etrntr-mom} shows the momentum dependence of the mean
total TR energy for the four radiators. In case of the {\it sandwich}
radiator this observable can also be extracted from two measurements
of total energy loss spectra in the drift chamber without radiators
(only d$E$/d$x$) and with radiators (d$E$/d$x$ + TR). In this case
the energy spectra are averaged over four (2002) and three chambers
(2004). The direct TR measurements are in general well reproduced
by the simulations, but for the {\it Reg2}, {\it sandwich} and
{\it fiber} radiator for momenta larger than 6\,GeV/c the measured
data points diverge from the saturated behavior predicted by the
simulations. This is probably a systematic error connected with an
underestimation of the SR background for higher momenta (Compare to
Fig. \ref{sr-mom}). For the indirect measurement one expects a larger
value for the total TR energy, due to the absorption of part of the
TR in the helium pipe, etc. This is found also in the data at momenta
up to 3\,GeV/c, but at higher momenta the data shows no saturation
and diverges from the prediction. The reason for this discrepancy is
not understood in detail yet, but it certainly is of importance for
a thorough understanding of the pion efficiencies achieved with the
ALICE TRD. An observed layer dependence of this effect
points at TR buildup and/or bremsstrahlung from the different
detector layers.

\subsection{Scaling with Radiation Length}

\begin{figure}[t]
  \begin{center}
    \includegraphics[width=8cm]{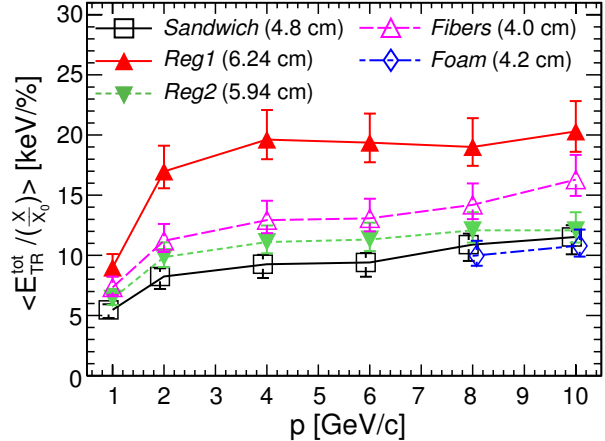}
    \caption{Momentum dependence of the total TR yield of the
     different radiators scaled with their radiation length $X/X_0$.}
    \label{etrntr-mom-x0}
   \hspace{5mm}
  \end{center}
\end{figure}

In this section we investigate the mean total observed TR energy for
the different radiators scaled with their radiation length:
$\langle E_{TR}^{tot}\rangle/\frac{X}{X_0}$. The radiation length is
a scaling variable for the probability of occurrence of bremsstrahlung
and pair production, and for the variance of the angle of multiple
scattering in the detector material. Since many transition radiation
detectors are also used in experiments where track reconstruction of
the particles is essential, a good TR performance should not be
achieved at the expense of larger material budget. As we can see from
Fig. \ref{etrntr-mom-x0}, the good TR performance of the thicker
regular radiator {\it Reg2} is achieved at the expense of about twice
the radiation length; this is the reason why in this context the
performance of {\it Reg2} is surpassed by {\it Reg1}. The
{\it sandwich} radiator is not performing as well as the other
radiators, which is due to the additional components used in its
construction (carbon fiber coating, epoxy and aluminized mylar foil).
In the case of the ALICE TRD, the radiators also serve the purpose
of reinforcing the large area detectors against the gas overpressure,
the weight of the Xe gas and the forces due to the sum of wire
tensions. Due to the low radiation length, pure fibers -- as used
in ATLAS\,\cite{ATLAS} -- should be considered an attractive
solution for tracking TRDs.

\section{Summary and Conclusions}

We have measured energy spectra of transition radiation (TR) for two
regular foil and some irregular radiators ({\it fiber}, {\it foam},
{\it sandwich}) for electrons in the momentum range from 1 to 10\,GeV/c
with drift chambers operated with Xe, CO$_2$(15\%). We used prototypes of
the drift chambers to be used in the Transition Radiation Detector (TRD)
of the ALICE experiment at CERN.

We extracted cluster number distributions and found that they can be
approximated by Poisson distributions. We observe a systematic increase
of the measured photon yield as a function of momentum. This can be
explained by synchrotron radiation, which is created by electrons in
the magnetic field used to deflect the beam. The assumption that the
increase of this photon background with momentum is due to synchrotron
radiation is well explained by simulations. After subtracting this
background, the mean values of the number of detected photons and of
the total TR energy are essentially constant above 3\,GeV/c.

The measured spectral shapes are quantitatively reproduced by
simulations, in particular for the two regular foil radiators.
For the {\it sandwich} and {\it fiber} radiators the simulation
parameters only reflect typical dimensions of the radiator
materials; we chose values that reproduce our data. With this
method the momentum dependence of the mean TR energy deposit is
quantitatively reproduced.

Our results significantly enhance the quantitative understanding
of transition radiation; this is in particular the case for the
performance of the {\it sandwich} radiators which are adopted in
the ALICE TRD.

\section*{Acknowledgements}

We acknowledge J. Hehner and A. Radu for their skills and dedication in
building our detectors, R. Glasow and W. Verhoeven for the supply of the
different radiators and N. Kurz for help on the data acquisition. We would
also like to acknowledge A. Przybyla for his technical assistance during the
measurements, I. Enculescu and the material science department at GSI for
the SEM photographs, P. Martinengo for help during setup and running
at CERN and the CERN PS personnel for their assistance.



\end{document}